\pdfoutput=1
\RequirePackage{ifpdf}

\documentclass[12pt,cits,hyper]{JINST}

\usepackage{booktabs}
\usepackage{setspace}

\newcommand{\degr}{\ensuremath{^{\circ}}}

\newcommand{\minitab}[2][l]{\begin{tabular}{#1}#2\end{tabular}}

\title{First Demonstration of Imaging Cosmic Muons in a Two-Phase Liquid Argon TPC using an EMCCD Camera and a THGEM}
\author{K.~Mavrokoridis\thanks{Corresponding author}, J.~Carroll, K.~J.~McCormick, P.~Paudyal, A.~Roberts, N.~A.~Smith, C.~Touramanis\\
University of Liverpool,
Department of Physics, Oliver Lodge Bld, Oxford Street, Liverpool, L69~7ZE, UK\\

E-mail: \email{k.mavrokoridis@liverpool.ac.uk}}

\abstract{

Colossal two-phase Liquid Argon Time Projection Chambers (LAr TPCs) are a proposed option for future long-baseline neutrino experiments.
This study illustrates the feasibility of using an EMCCD camera to capture light induced by single cosmic events 
in a two-phase LAr TPC employing a THGEM. An Andor iXon Ultra 897 EMCCD camera was externally mounted via
a borosilicate glass viewport on the Liverpool two-phase LAr TPC. 
The camera successfully captured the secondary scintillation light produced at the THGEM holes that had been induced by cosmic events.
The light collection capability of the camera for various EMCCD gains was assessed.  For a THGEM gain of 64 and an EMCCD gain of 1000, 
clear images were captured with an average signal-to-noise ratio of 6.
Preliminary 3D reconstruction 
of straight cosmic muon tracks has been performed by combining the camera images, PMT signals and THGEM charge data. 
Reconstructed cosmic muon tracks were used to determine THGEM gain and to calibrate the intensity levels 
of the EMCCD image.

}


\begin{document}

\section{Introduction}

Realisation of colossal-scale Liquid Argon~(LAr) Time Projection Chambers~(TPCs) for long-baseline neutrino physics is underway, with international efforts
such as the DUNE collaboration bringing together the single and two-phase LAr TPC technologies~\cite{Rubbia:2004p3272, Rubbia:2009p3284, LBNE, DUNE}. A large scale two-phase LAr TPC demonstrator is currently planned at CERN to support the LBNO-DEMO experiment~\cite{Agostino:2014}. Small scale two-phase LAr TPCs have demonstrated 
excellent calorimetric and tracking capabilities using a charge readout approach with THGEMs~\cite{Badertscher:3litre, Badertscher:2012}. This paper is a continuation of our previous publication, reporting on our new optical readout approach in a two-phase LAr TPC 
using THGEMs and a camera system~\cite{kostasccd}.

Two-phase LAr TPCs detect the prompt scintillation light~(S1) produced in the liquid phase and free-ionised electrons, generated during an interaction, are drifted to the liquid surface where they are extracted to the gas phase producing secondary delayed electroluminescence light~(S2). 
These electrons then enter the THGEM holes and are amplified by
the high electric field, creating further secondary scintillation light production as a result of electroluminescence. With application of a 
high enough electric field across the THGEM holes, avalanche is induced and the electroluminescence produced increases exponentially with THGEM electric field~\cite{Monteiro:2007, Monteiro:2008, Phil:2009, Monteiro:2012}. In the case of the proposed future giant scale two-phase LAr TPCs the amplified electrons are drifted to a segmented anode plane for charge readout.

Whilst THGEMs provide excellent gain resulting in good signal-to-noise ratios, the current THGEM charge readout approach has challenges with respect to scale-up: separate readout channels are required for every anode strip, which can number in the order of up to a million for the colossal scale detectors, and the resolution of
tracking reconstruction is limited by the pitch of the strips.  An alternative readout method could involve the exploitation of the secondary scintillation light produced in the THGEM holes~\cite{Phil:2009, Bondar:2010, Bondar:2012A, Bondar:2012B}. 

In our previous publication we reported the successful demonstration of the feasibility of capturing the secondary scintillation light produced by the 
THGEM using a CCD camera internally mounted within the cryogenic environment of a two-phase TPC. For the first time this technology was used to image 
the secondary scintillation light induced by an internal Am-241 alpha source and an external Cs-137 gamma source in an integrated time window~\cite{kostasccd}. 
Imaging of secondary scintillation light with a CCD camera has many advantages over the current charge readout based approaches
such as ease of customisation and upgrade, cost effectiveness and elegant simplicity.

We have now utilised an upgraded system
employing an externally mounted Electron Multiplying Charge Coupled Device~(EMCCD) camera and we have demonstrated for the first time the calorimetric and tracking reconstruction capabilities of this technology for single
cosmic events.

\begin{figure}[t]
\begin{center}
\begin{tabular}{c}
	\includegraphics[width=.65\textwidth]{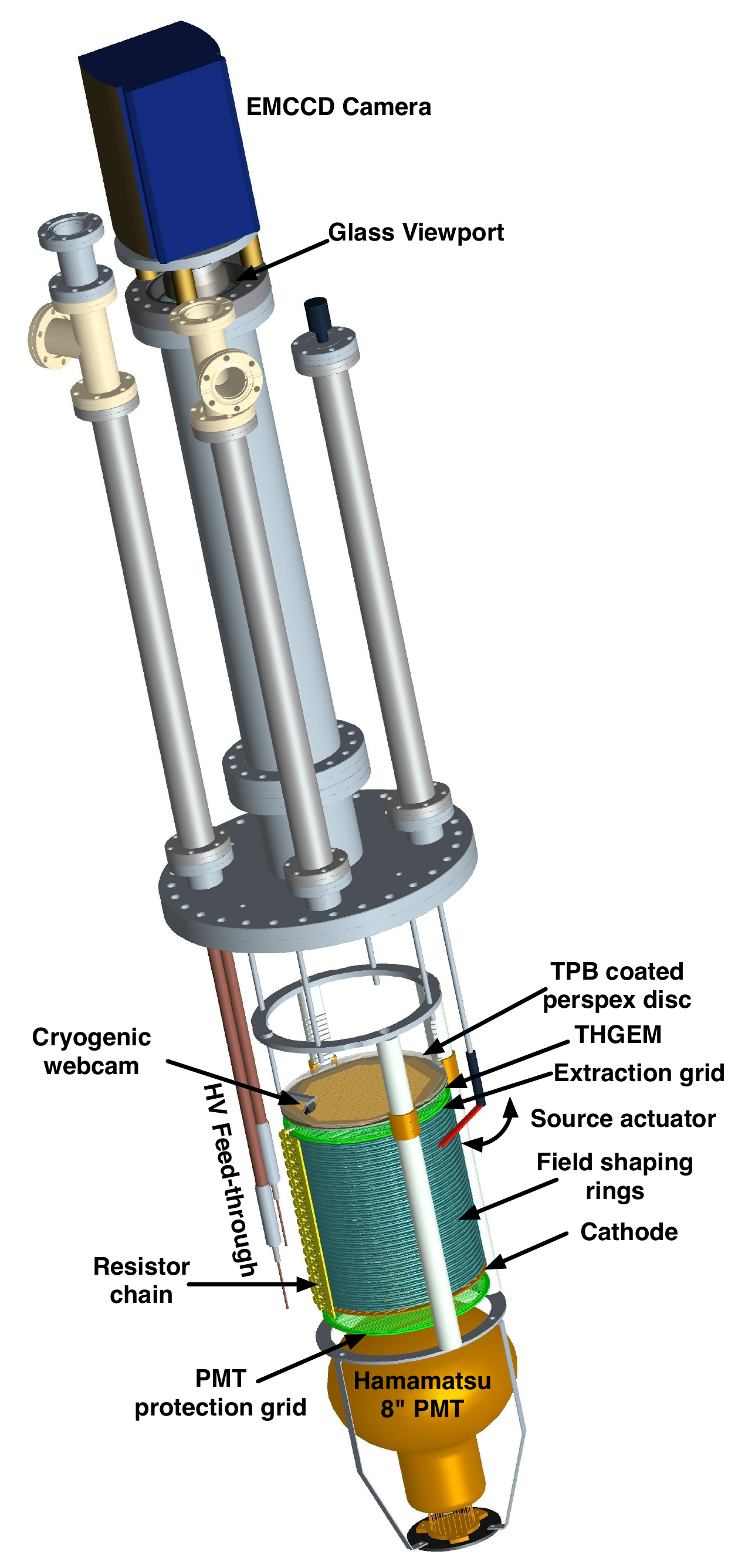}
\end{tabular}
\end{center}

\caption{A 3D CAD model of the detector assembly.}

\label{model}
\end{figure}

\section{The Liverpool LAr Setup} \label{setup}

A CAD model of the internal detector assembly is shown in Figure~\ref{model}. 
An 8-inch Hamamatsu R5912-02MOD PMT faces upwards towards the 19~cm long field cage above which the electron extraction grid and THGEM are positioned.
An Andor iXon Ultra 897 EMCCD camera utilising a 50 mm f/1.2 Nikon NIKKOR lens was placed externally at ambient temperature 
on a borosilicate viewport approximately 1~m above and focused on the THGEM plane. 

The PMT, the 3M{\small\texttrademark}-foil reflector surrounding the field cage (the reflector is not shown in Figure~\ref{model}) and the bottom face of a perspex disk positioned above the THGEM plane
are coated with tetraphenyl-butadiene (TPB) in order to shift the 
argon Vacuum Ultra Violet~(VUV) scintillation light to 430~nm, well within the PMT and EMCCD high quantum efficiency range~\cite{WLSkostas}.
The 150~cm$^2$ THGEM, which is in the gaseous phase, is 1mm thick and the $\approx$23000 holes have a diameter of 500~$\mu$m and a 50~$\mu$m dielectric rim is etched 
around each hole to extend the breakdown voltage of the THGEM. This is the same type of THGEM that was used in our previous study.
The PMT and the THGEM charge pre-amp~(ORTEC 142IH) signals are recorded using an Agilent DP1400 digitiser.
A level monitor cryogenic webcam is positioned to allow monitoring of the extraction region. A more detailed description of the detector subcomponents can be found in~\cite{kostasccd}.

The internal detector assembly is suspended from a 13.25-inch conflat (CF) flange into a 40~litre Ultra High Vacuum (UHV) chamber. It is through this flange that
all electrical and gas feedthroughs are mounted, as well as the EMCCD camera viewport. A LAr recirculation/purification system is located on the side of the 40~litre chamber.
The chamber sits within a vacuum jacketed open 250~litre LAr bath to maintain the cryogenic temperature of the detector.
Further details of the cryogenic system and the novel recirculation/purification system are given in our previous publication~\cite{kostaspurity}.

\subsection{Characteristics of the EMCCD Camera}

The Andor iXon Ultra 897 EMCCD camera utilises a 512 $\times$ 512 back illuminated e2v sensor with high pixel well depth capacity and a high quantum 
efficiency in the region of 80~\% at 430~nm. 
Advantageously over conventional CCD cameras, the EMCCD camera features an electron multiplication capability with a gain of up to 1000,
allowing for an effective readout noise of less than 1 electron therefore making the camera single photon sensitive. At full resolution the camera can capture 56 frames/sec,
for 4$\times$4 binning 206 frames/sec can be recorded. In crop mode (at 512$\times$1) the camera can record 11,074 frames/sec.
Table~\ref{ccd_manufacture} summarises the key specifications of the camera system. For further details of the camera system 
see~\cite{Andor}.

\begin{table}
\caption[Andor EMCCD Characteristics.] {Andor iXon Ultra 897 EMCCD Characteristics.}

\begin{center}
\vspace{4mm}
{\footnotesize \begin{tabular}{lc}
\toprule
\multicolumn{2}{c}{Andor iXon Ultra 897 EMCCD} \\
\midrule
CCD sensor type: & e2v \\ 
CCD sensor QE option: & Back illuminated, dual AR coated\\
Image area: & 8.2~mm$\times$ 8.2~mm \\
Pixel resolution (H$\times$V): & 512$\times$512, 0.262 Megapixels \\
Pixel size: &16~$\mu$m~$\times~$16~$\mu$m \\
Active area pixel well depth: & 180,000  e$^-$\\
Gain register pixel well depth: & 800,000 e$^-$\\
Linear absolute Electron Multiplier gain: & 1- 1000\\
Read out noise:  & <~1~e$^-$ with electron multiplication \\
Spectrum range: & 300~nm - 1050~nm \\
Quantum efficiency at 430~nm: & ~80 \% \\
Vertical clock speed: & 0.3~$\mu$s to 3.3 $\mu$s (variable) \\
Frame Rate: & 56 - 11,074 fps \\
Digitisation: & 16 bit \\
Triggering: & Internal, External, External Start, External Exposure, Software Trigger \\
Data interface: & USB 2.0 \\
\bottomrule

\end{tabular}}
\end{center}
\label{ccd_manufacture}
\end{table}

\section{Experimental Procedure}

\subsection{Operation Principle}

When a particle passes through the detector medium an ionisation trail and prompt scintillation light (S1) are produced. 
The prompt scintillation light is immediately collected by the PMT signalling the start of the event, 
and the free ionisation electrons are drifted within the uniform electric field to the liquid surface where a higher electric field is applied to extract the electrons to the gas phase. The electrons are then accelerated within the THGEM holes 
creating avalanche and further secondary scintillation light (S2).
This secondary scintillation light is seen by the PMT at the bottom and by the EMCCD camera at the top.

\subsection{Detector Preparation}
To prepare the detector for operation it is evacuated to $1\times10^{-7}$~mb and filled with LAr via liquefaction and filtration. The LAr level is adjusted to
halfway between the extraction grid and THGEM. The LAr is then purified further utilising a custom made LAr recirculation/purification system. For 
further details on the operation and preparation of the detector see~\cite{kostasccd}.

\subsection{Electric Field Configuration}
The electric field configurations applied are summarised in Table~\ref{LAr_Efields}.
A uniform electric field of 500~V/cm was established by applying 18~kV to the cathode and 8~kV to the extraction grid.
For the 19 cm drift length and 0.5 kV/cm electric field, the electrons will take $\approx$120~$\mu$s to reach the liquid surface~\cite{Walkowiak:2000}.  

\subsection{EMCCD Operation}
The EMCCD sensor was maintained at -80~$\degr$C utilising the built-in thermoelectric cooling system. For the presented data, 4$\times$4 binning and an electron
multiplication gain of 1000 was utilised, however the camera performance was also evaluated for various gains and binning configurations. The camera exposure was set to 1~msec to contain full track information, and exposures down to 10~$\mu$sec were assessed.


\begin{table}
\caption[Electric Field Configuration] {Configuration of the electric fields applied in two-phase operation.}
\begin{minipage}{\textwidth}
\begin{center}
\vspace{4mm}
{\footnotesize \begin{tabular}{lcccc}
\toprule
 & \minitab[c]{Distance to the \\ stage above (cm)} 
& Potential (kV) & \minitab[c]{Field to the stage \\ above (kV/cm)}\\
\midrule
THGEM (top electrode) & - & +1.4 & - \\
THGEM (bottom electrode) & 0.1 & -2.3 to -2.6 & 37 to 40  \\
Extraction grid & 1.0 & -8  &
 \minitab[c]{8.10 to 8.55 (gas)\footnote{The field in gas is 1.5 times stronger than in liquid due to the change of the dielectric constant.}\\5.40 to 5.70 (liquid)} \\
Cathode & 19 & -18  & 0.52  \\
\bottomrule
\end{tabular}}
\end{center}
\end{minipage}
\label{LAr_Efields}
\end{table}

\subsection{Trigger Configuration}
The camera was triggered externally by the PMT primary scintillation signal.  
For each event, data were recorded simultaneously by the PMT, the charge preamp and the camera for a 1~msec time window. 
A TTL trigger was provided to the camera, and a NIM trigger to the PMT/THGEM digitiser (Agilent DP1400).
The maximum trigger rate delivered to the camera and the digitiser was less than $\approx$ 40~Hz in order to allow for enough time for each frame 
to be read out.
This was achieved by introducing a veto to the trigger output, such that following a trigger, another trigger cannot be sent within a specified time window, thus limiting the trigger rate to the required 40~Hz.

\subsection{Data Format}
The EMCCD data were written in an Andor native sif format and then converted to ascii and also FITS format taking advantage of the many available image analysis 
tools developed by the astronomy community.
The PMT and charge data were written in ascii format. A ROOT~\cite{ROOT} script was developed to bring together the analysis of the three data outputs. 

\section{Results \& Analysis}

\subsection{Muon Imaging}


\begin{figure}[t!]
\begin{center}
\begin{tabular}{c}
	\includegraphics[width=1.\textwidth]{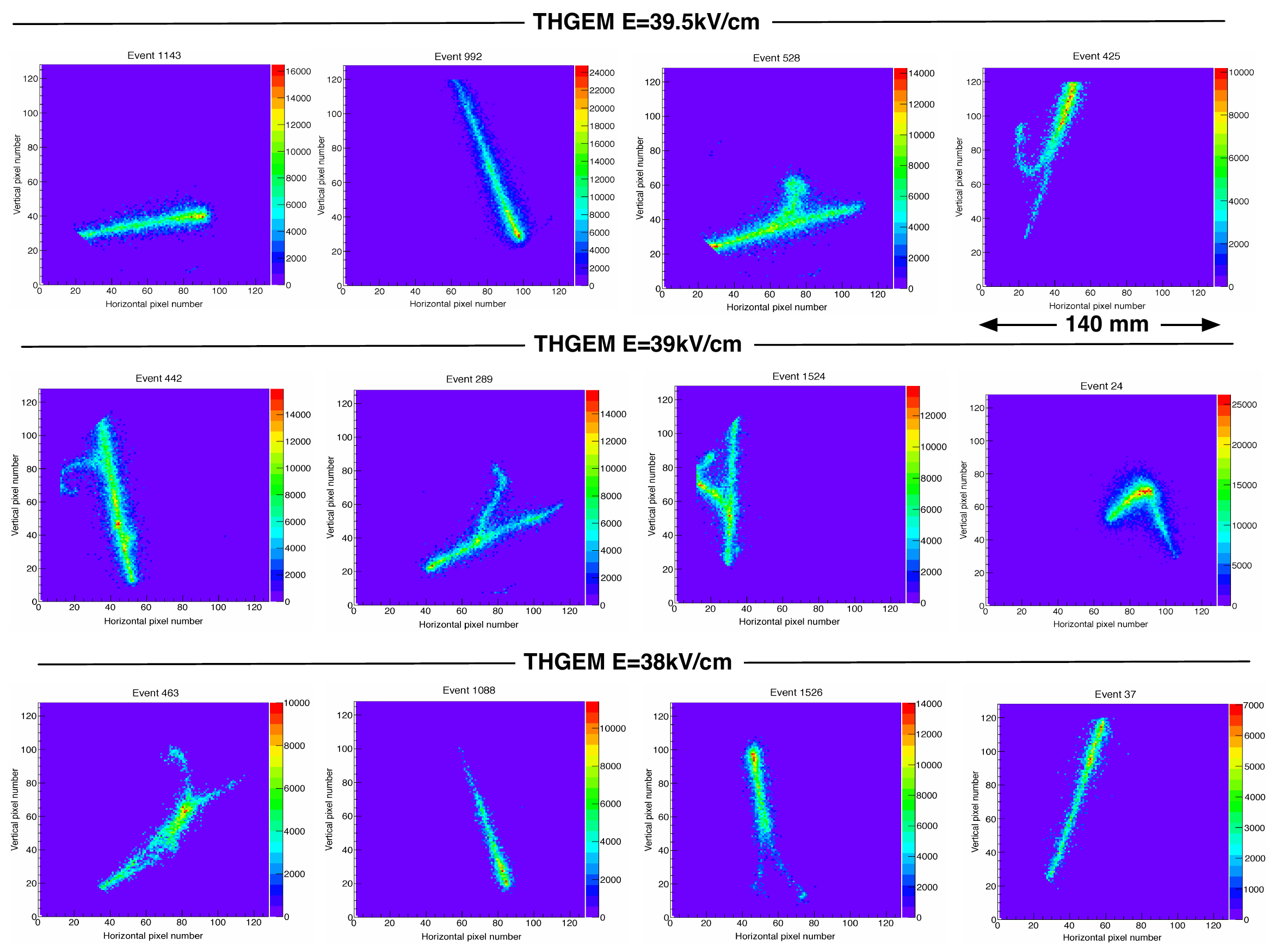}
\end{tabular}
\end{center}

\caption[Cosmic_gallery]
{ A gallery of cosmic events passing through the detector medium for various THGEM electric fields. The events were captured using 4$\times$4 binning, 1~msec exposure and an EMCCD of gain 1000. One pixel corresponds to 1.1~mm.}

\label{eventsample}
\end{figure}

\begin{figure}[t!]
\begin{center}
\begin{tabular}{c}
	\includegraphics[width=0.7\textwidth]{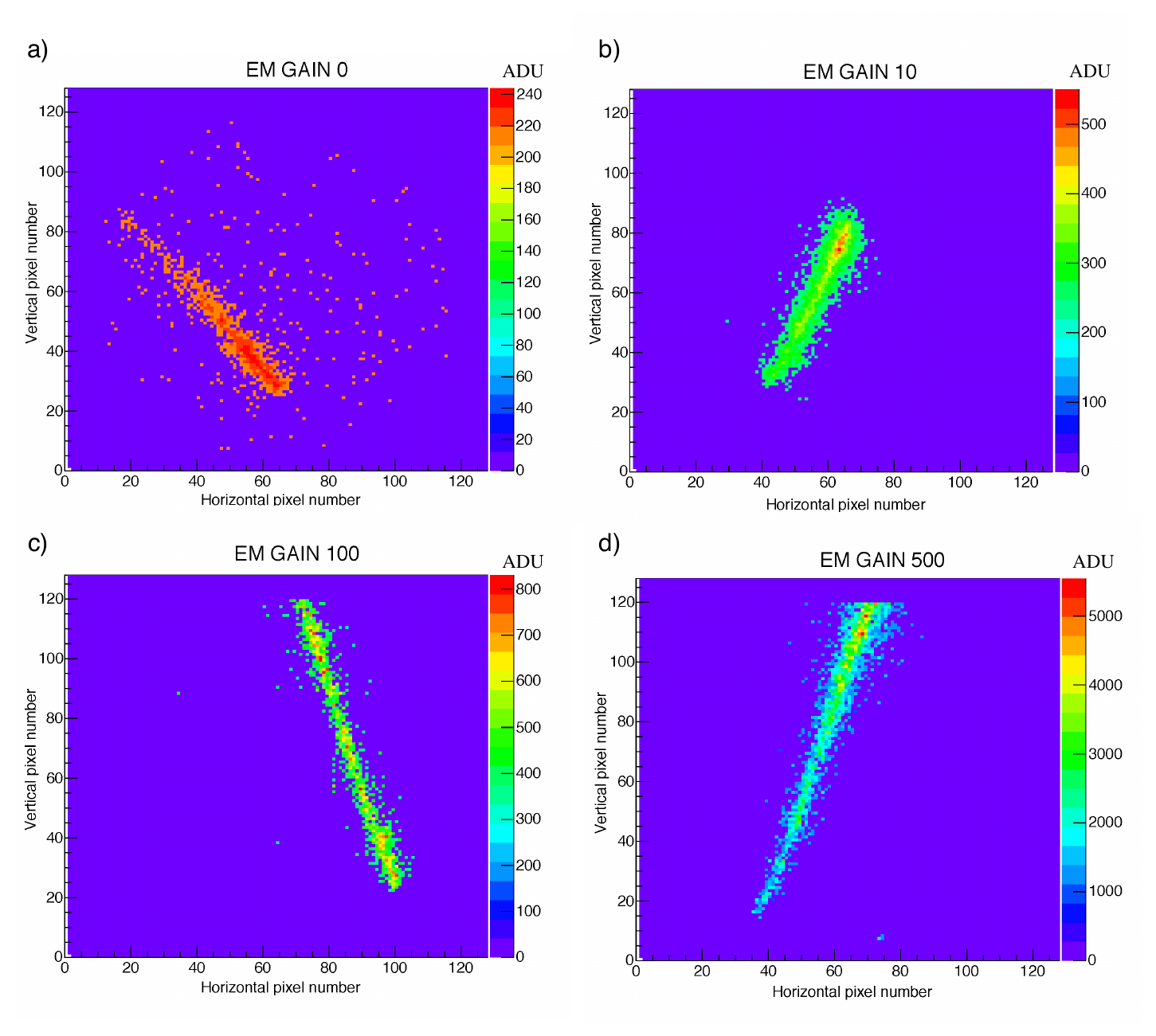}
\end{tabular}
\end{center}

\caption[Gain_Galery]
{ A comparison of images of cosmic events for EMCCD gain a)  0, b) 10, c) 100 and d) 500.
The camera binning was 4$\times$4, the exposure was 1~msec  and the THGEM field was constant at 39~kV/cm.}
\label{gain_galery}
\end{figure}

The EMCCD camera successfully captured the secondary scintillation light produced in the THGEM holes induced by cosmic muon events. In terms of both resolution and
signal-to-noise performance the best images were obtained using 4$\times$4 binning and an EM gain of 1000. 
A gallery of cosmic events captured with these camera settings, where 1~pixel corresponds to 1.1~mm,
and three different THGEM electric fields (THGEM gains 64, 41 and 19) are presented in Figure~\ref{eventsample}. It is also notable that the camera
was able to capture light at 1$\times$1 binning using an EM gain of 1000 and at 4$\times$4 binning the signal was visible even without any camera gain. For comparison,
a sample of events taken at 0, 10, 100 and 500 EMCCD gain are shown in Figure~\ref{gain_galery}.


\subsection{Signal-to-Noise Ratio}
\label{title:SNR}
The mean background pixel intensity level for 1~msec exposure and an EM gain of 1000
was 163~ADU~(Analogue Digital Unit) and the dark shot noise, which can be considered to be  the standard deviation of the background pixel intensity levels,
was approximately 20~ADU. The dark shot noise is only significant for dark or extremely low intensity signal images. 
As the images had high intensity levels (at EM gain 1000), for the purpose of this
paper we have only considered the shot noise ($\sqrt {Pixel~Intensity~(e^{-})}$), where the pixel intensity is measured in electrons,  and the EMCCD noise factor~($\sqrt {2}$)
occurring from the random nature of the charge amplification process, as the major contributors to the noise~\cite{andorSNR}.
Thus the signal-to-noise for any pixel can be represented by:
\begin{equation} 
SNR = \frac{Pixel~Intensity~(e^{-})}{\sqrt {2\times Pixel~Intensity~(e^{-})}}
\label{SNR}
\end{equation}
A signal is considered observable when the SNR~$>$1. By equating eqn~\ref{SNR} to 1, a minimum detectable signal of 2e$^{-}$ can be determined. 
When using the calibration tools provided by the manufacturer within the Andor Solis software, for an EM gain of 1000 
a signal of 2e$^{-}$ per pixel corresponds to 400 ADU. Thus, any pixel intensity above 400 ADU is a significant observation.

\subsection{3D Reconstruction}

\begin{figure}[t]                         
\begin{center}
\begin{tabular}{c}
	\includegraphics[width=0.9\textwidth]{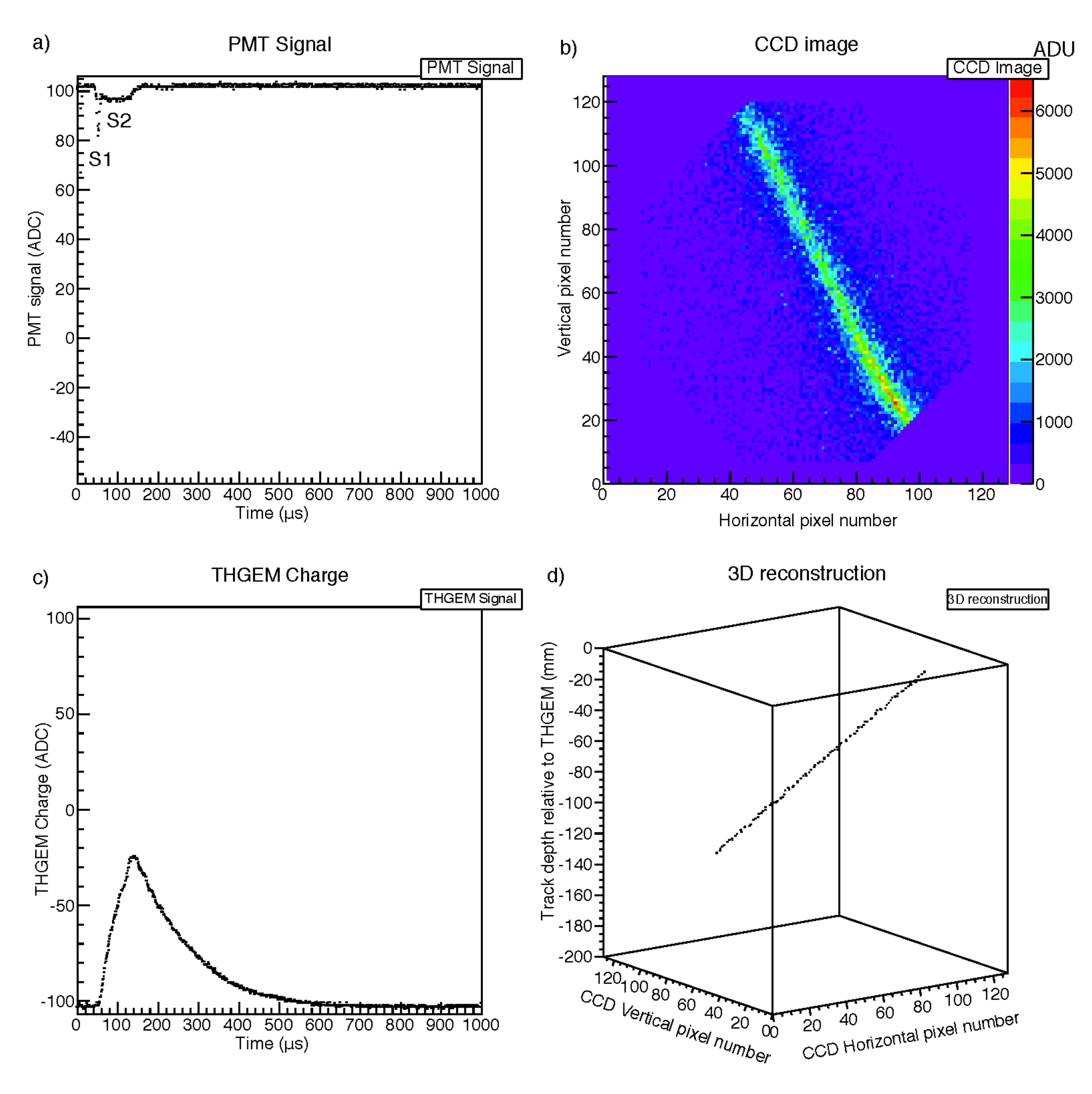}  
\end{tabular}
\end{center}

\caption{a) The primary and secondary scintillation signal seen by the PMT as a cosmic muon passes through the detector medium; b) The same secondary scintillation signal captured by the EMCCD camera (raw data); c) The overall charge collected by the THGEM; d) A 3D reconstruction of the cosmic muon by correlating the PMT data, THGEM charge and EMCCD photograph.}

\label{3Dreconstuction}
\end{figure}

For THGEM electric fields between 36~kV/cm to 39.5~kV/cm, 2000 muon triggered images were collected and correlated to the PMT and THGEM charge preamp data. 
The interval between taking measurements at each electric field was kept to $\approx$30 minutes thus minimising possible concerns of THGEM gain stability which has been shown to decrease over a day of operation before reaching an equilibrium~\cite{Cantini:2015}.
By combining the PMT prompt (S1) and secondary (S2) scintillation light signals, the THGEM charge signal and the EMCCD image data a preliminary 3D reconstruction algorithm for straight line tracks was developed. The depth of the interaction was determined from the time difference between the secondary and prompt scintillation pulses seen by the PMT. The x, y co-ordinates were known from the EMCCD image; a linear 2D fit on the EMCCD image provides a plane in the detector volume where the track has occurred. 
The path length (d$x$) was determined by:  

\begin{equation}
\mathrm{d}x= \sqrt {(x-x_{0})^{2} + (y-y_{0})^{2} + (z-z_{0})^{2}}
\label{dx}
\end{equation}

where $z_0$ is the delay between the S1 signal and the start of the charge rise time and $z$ is the charge rise time (converted to mm, 120~$\mu$s e$^-$ drift time $\approx$190~mm at 0.5~kV/cm) plus $z_0$.
A 3D event reconstruction sample is shown in Figure~\ref{3Dreconstuction}.

\subsection{Energy Calibration}

Cosmic muons were used to determine the THGEM gain and to calibrate the intensity levels of the EMCCD image to keV.  

\begin{figure}[t]
\begin{center}
\begin{tabular}{c}
	\includegraphics[width=1.0\textwidth]{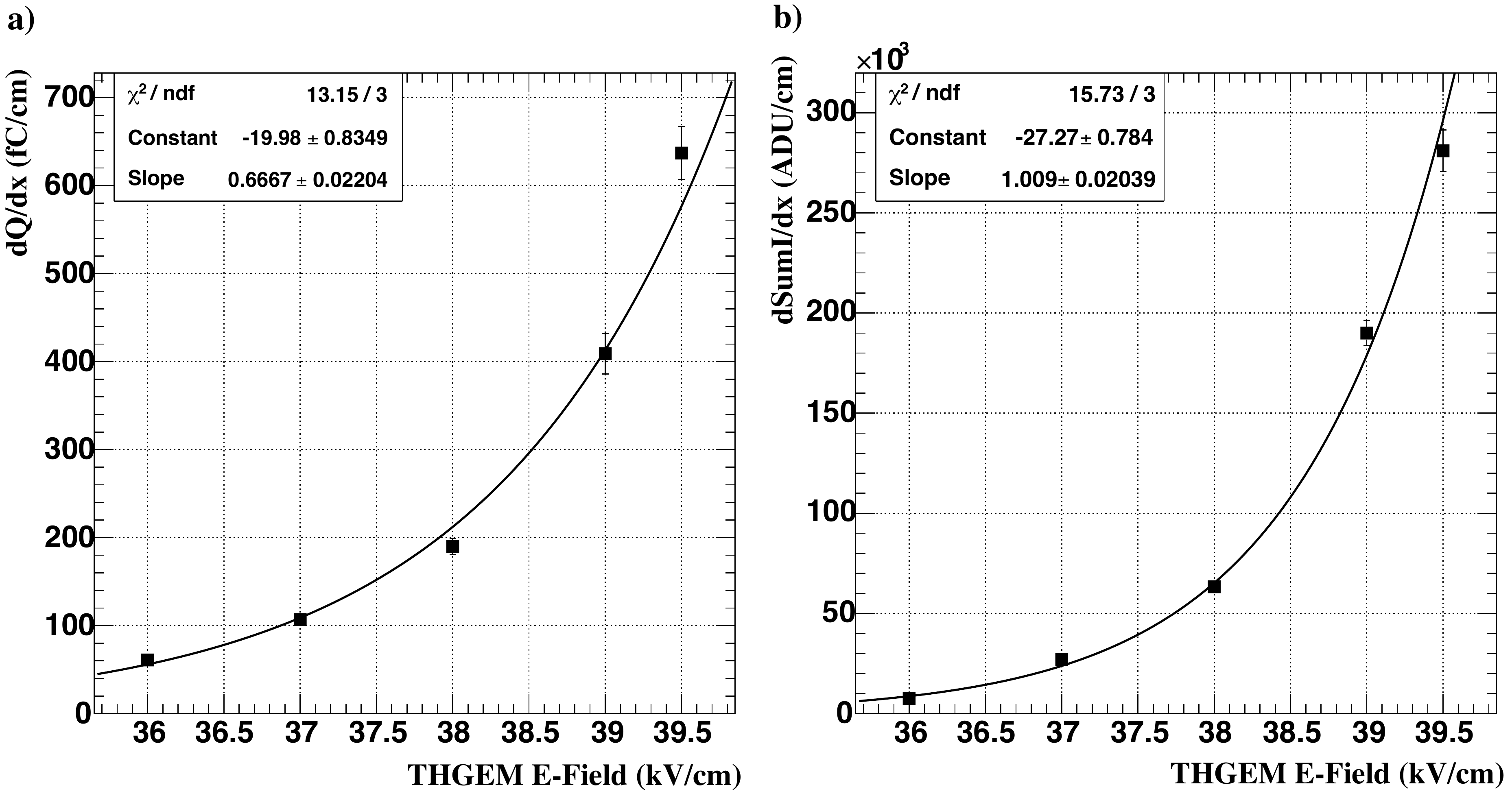}
\end{tabular}
\end{center}

\caption{a) Charge loss per unit track length for various THGEM electric fields. b) Light loss per unit track length as captured by the camera for various THGEM electric fields. For the highest E-Field value the THGEM gain corresponds to approximately 64 (Using 10~fC/cm muon charge deposition and no corrections for additional charge losses such as grid transparency).}

\label{chargelossvsEfield}
\end{figure}

The charge loss per unit path length (d$q$/d$x$) of the cosmic muons was established and the MPV values of the d$q$/d$x$ Landau distribution for each 
THGEM electric field measurement is presented in Figure~\ref{chargelossvsEfield}a.
Since the charge deposition of muons in liquid argon is 10~fC/cm (taking into account the 30\% recombination factor at 0.5~kV/cm drift field, 
but with no further correction for electron transparency of the extraction grid) the gain of the THGEM was established for each applied electric field up to
39.5~kV/cm, which had a gain of 64.
The maximum electric field of the THGEM before breakdown was determined to be just above 40~kV/cm. 

\begin{figure}[t!]
\begin{center}
\begin{tabular}{c}
	\includegraphics[width=.8\textwidth]{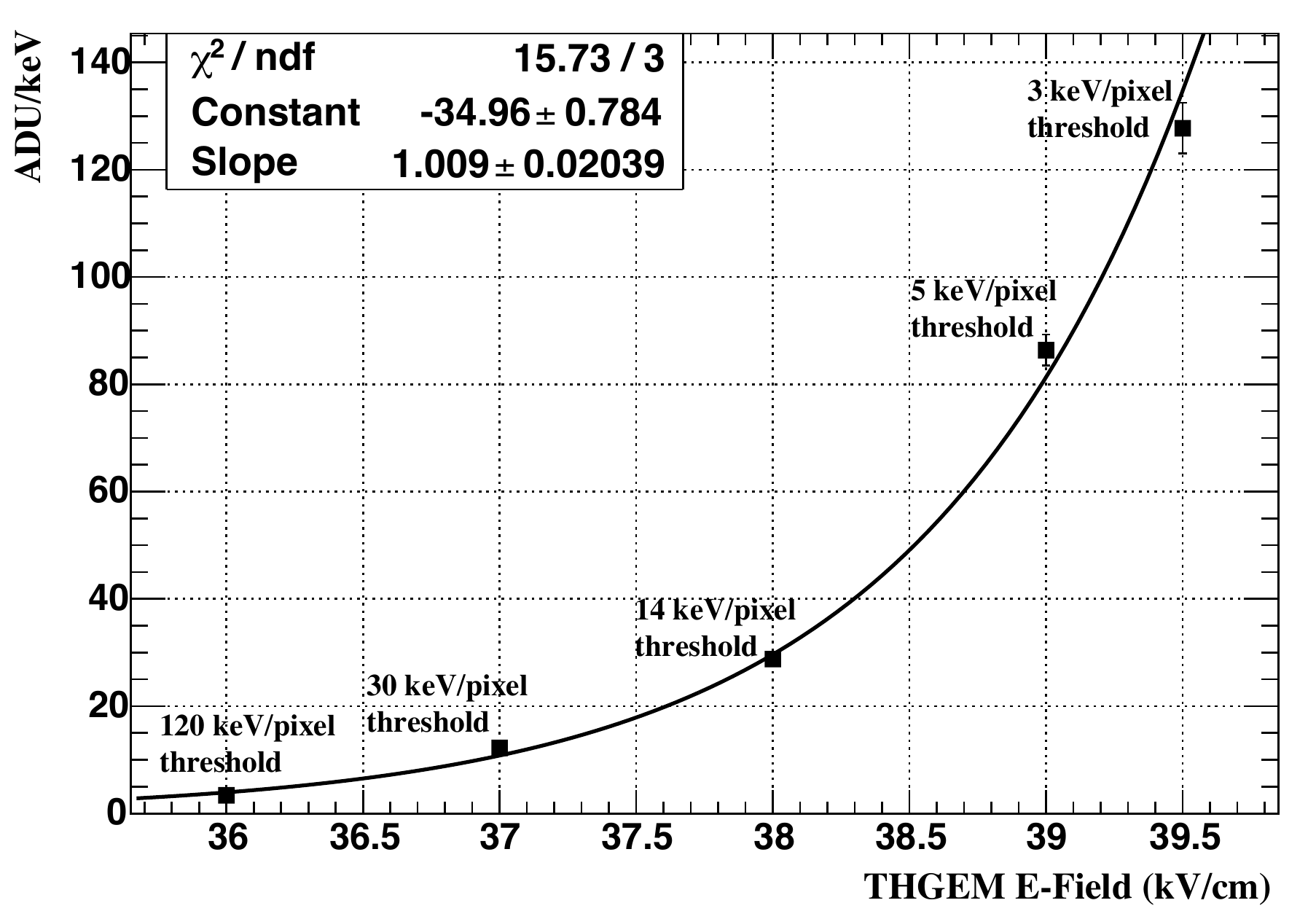}
\end{tabular}
\end{center}

\caption{ Energy calibration using cosmic muon energy deposition of  pixel intensity levels (ADU) to keV for various THGEM E-Fields. 
The energy threshold  per pixel for each measurement point is also depicted by taking into account the minimum observable signal of 400~ADU at EM gain 1000.} 

\label{fig:Energy_calibration}
\end{figure}

Similarly, the light loss per unit path length (d$sumI$/d$x$) seen by the camera for each identified muon track is shown in Figure~\ref{chargelossvsEfield}b.
Both d$sumI$/d$x$ and d$q$/d$x$ follow a similar exponential relationship with an increase in THGEM electric field. 
Using the expected energy deposition of muons in LAr ($\approx$~2.2 MeV/cm~\cite{ArgoNeuT:2012}), the image intensity levels were calibrated from ADU to keV. 
Figure~\ref{fig:Energy_calibration} shows the calibration of intensity levels to keV for the various THGEM electric fields. 
 At a THGEM electric field of 39.5~kV/cm and camera gain 1000, 128~ADU corresponds to 1~keV.
For the highest THGEM gain of 64, the pixel energy threshold for muon events was determined to be $\approx$~3~keV by taking into account the minimum observable signal of 400~ADU as described in Section~\ref{title:SNR}. 

When an EM gain of 1000 and 4$\times$4 binning was used, the camera was demonstrated to detect individual cosmic events for a THGEM gain as low as 6.

For muon tracks, the average signal-to-noise at a THGEM gain of 64 was calculated to be $\approx$~6 whereas at a THGEM gain of 6, the average
signal-to-noise was $\approx$~3.


\section{Discussion \& Conclusions}

We have successfully demonstrated that the amount of secondary scintillation light produced in the THGEM holes and the sensitivity of an
 Andor iXon Ultra 897 EMCCD camera are compatible for allowing the imaging of cosmic events. At a THGEM gain of 64 and camera settings of 4$\times$4 binning and a camera gain of 1000
the signal could be distinguished very clearly with an average signal to noise ratio for cosmic events of 6.  

By correlating and combining the EMCCD image, PMT signal and THGEM charge from the events, successful 3D reconstruction of straight tracks was performed. In order to perform 3D reconstruction
without the aid of the THGEM charge readout, i.e. using solely the image data, the camera speed would need to be 100,000 frames/sec.  This would allow for the
segmentation in time of an event every 10~$\mu$sec.  Presently scientific CMOS cameras that can reach up to 1 million frames/sec are available, however
due to their high noise they are lacking the sensitivity required for this application. For example, we have tested the Photron FASTCAM Mini UX100~\cite{Photron}
which can capture up to 800,000 frames/sec with no successful results. However, CMOS technologies are advancing very quickly and are likely to 
improve in terms of quantum efficiency and readout noise in coming years. 
With current technologies, further light collection and spatial resolution improvements 
could be achieved using faster lenses (f/0.95) and an EMCCD camera with a larger sensor area such as the Andor iXon Ultra 888.
A larger ton-scale prototype of our presented set-up with a 1~m$^{2}$ THGEM assembly plane
would require 9 EMCCD Andor iXon Ultra 888 cameras and segmentation of the THGEM assembly into pads (the number of which would depend on the resolution required on z) for charge readout.

Optical readout presents many clear advantages when compared to currently accepted segmented THGEM charge readout technologies. Optical readout is scalable, without introducing the complexity and cost required by the vast segmented charge readout anode planes placed above the THGEMs. Mounted externally from the TPC, the EMCCD cameras are easily accessible, simplifying detector maintenance and allowing for uncomplicated installation of detector upgrades as camera technology develops, whereas upgrade of currently proposed segmented anode planes would require TPC disassembly, introducing significant cost and time expenditure. 
The range of settings available with camera technology, such as binning, readout speed and exposure time, allows for detector readout customisation on the fly and customisation to suit the needs of a range of different experiments
from precision track reconstruction using full camera resolution to the search for low energy interactions at very high camera binning
using the same TPC.

We therefore propose that EMCCD cameras could be employed in future giant scale two-phase LAr neutrino experiments as an alternative readout method, replacing the highly segmented anode planes.

\acknowledgments

The Authors are grateful for the expertise and dedicated contributions of the Mechanical Workshop of the Physics
Department, University of Liverpool. We also thank John Bland from the HEP IT group for his support.
We acknowledge the support of the University of Liverpool and STFC.



\end{document}